\documentstyle[aps,epsfig]{revtex}
\begin{document}
\draft
\tighten
\title{Enhancement of gluonic dissociation of $J/\psi$ in viscous
QGP}

\author{\bf A. K. Chaudhuri\cite{byline}}
\address{ Variable Energy Cyclotron Centre\\
1/AF,Bidhan Nagar, Calcutta - 700 064\\}
\date{\today}

\maketitle

\begin{abstract}
We  have  investigated  the  effect  of  viscosity on the gluonic
dissociation of $J/\psi$ in an equilibrating plasma.  Suppression
of $J/\psi$ due to gluonic dissociation depend on the temperature
and  also on the chemical equilibration rate. In an equilibrating
plasma, viscosity affects the temperature evolution and also  the
chemical  equilibration  rate,  requiring  both of them to evolve
slowly compared to their ideal counter part. For Au+Au collisions
at RHIC  and  LHC  energies,  gluonic  dissociation  of  $J/\psi$
increases for a viscous plasma. Low $P_T$ $J/\psi$'s are found to
be  more  suppressed  due  to viscosity than the high $P_T$ ones.
Also the effect is more at LHC energy than at RHIC energy.
 \end{abstract}


In  relativistic  heavy  ion  collisions $J/\psi$ suppression has
been recognized as an important tool  to  identify  the  possible
phase  transition to quark-gluon plasma \cite{vo99,ge99}. Because
of the large mass of  the  charm  quarks,  $c\bar{c}$  pairs  are
produced  on  a  short  time scale. Their tight binding also make
them immune to final state interactions. Their  evolution  probes
the  state of matter in the early stage of the collisions. Matsui
and Satz \cite{ma86} predicted that in  presence  of  quark-gluon
plasma (QGP), due to color screening, binding of $c\bar{c}$ pairs
into  $J/\psi$  meson  will be hindered, leading to the so called
$J/\psi$ suppression in heavy ion collisions  .  Apart  from  the
color screening of $J/\psi$, gluonic dissociation of $J/\psi$ may
be  important  source  of  $J/\psi$  suppression  at RHIC and LHC
energies \cite{xu96,pa01}. Energy  dependence  of  gluon-$J/\psi$
inelastic  cross  section  shows  a  strong  peak  just above the
breakup  threshold  of  the  gluon  energy $\varepsilon_0 =2M_D -
M_{J\psi}$, where $M_{J/\psi}$ and $M_D$ are the $J/\psi$  and  D
meson masses \cite{kh94}. In the pre equilibrium stage, the parton
momenta will be high enough to break up a $J/\psi$. This break up
process will continue during the equilibration process, until the
temperature  drops  below  a  certain  value, or the beginning of
hadronisation.

Gluonic  dissociation  of  $J/\psi$  depends  on  the temperature
evolution and on the chemical equilibration rate \cite{xu96}.  It
is well known that dissipative effects like viscosity affects the
temperature  evolution  as well as chemical equilibration rate in
an equilibrating plasma, requiring both of them to evolve slowly,
compared  to  their   ideal   counterpart   \cite{ch00}.   As   a
consequence,    the    pre-equilibrium    stage    is   extended.
Correspondingly  gluonic  dissociation  of  $J/\psi$  should   be
increased due to viscosity, $J/\psi$'s will now have more time to
interact  with  a  gluon and get suppressed. In the present brief
report, we have investigated  the  effect  of  viscosity  on  the
gluonic dissociation of $J/\psi$. As will be shown below $J/\psi$
suppression  due  to  gluonic dissociation increases in a viscous
plasma. Suppression shows  strong  $p_T$  dependence,  low  $p_T$
$J/\psi$'s  being  more  suppressed  than  the  high  $p_T$ ones.
Suppression is also energy dependent, more at LHC energy than  at
RHIC energy.

We  assume  that  after a collision, a symmetric partonic system
(QGP) is formed in the central rapidity region.  We  also  assume
that  the  partonic system quickly achieve kinetic equilibrium by
time  $\tau_{iso}$  when  momenta  of  partons   become   locally
isotropic.  Beyond $\tau_{iso}$ further expansion of the partonic
system can be described by hydrodynamical  equations.  We  assume
that  the dominant reactions governing the chemical equilibration
process are the two body reactions $(gg\leftrightarrow  ggg$)
and  gluon  multiplication  and its inverse process, gluon fusion
($gg\leftrightarrow ggg$). The hot matter continues to expand and
cool due to  expansion  and  chemical  equilibration,  until  the
temperature fall below the critical value ($T_c$=160 MeV) at time
$\tau_f$.  We  assume that the hydrodynamical expansion is purely
longitudinal. As indicated in \cite{sr97}, at RHIC energies  the
transverse expansion effect is minimal. It does effect the parton
equilibration  rate at LHC energies, the effect showing sensitive
dependence on the initial condition of the  plasma.  For  initial
conditions  as  obtained from HIJUNG calculations, this effect is
not large \cite{sr97}

A $J/\psi$ produced at point ${\bf r}$ with  velocity  ${\bf  v}$
will travel a distance

\begin{equation}
d = - r cos \phi + \sqrt{R^2_A-r^2(1-cos^2 \phi)}
\end{equation}

\noindent  in  the time interval $t_\psi = M_T d /P_T$, before it
escapes from the partonic system of transverse  extension  $R_A$,
$\phi$  being  the  angle  between  the vectors ${\bf r}$ and
${\bf v}$. The total amount of time the $J/\psi$  remains  in  the
plasma   and  interact  with  a  gluon  is  the  smaller  one  of
$\tau_\psi$ and $\tau_f$, the life time of the plasma.

The  survival  probability  of  the  $J/\psi$  averaged  over its
initial position and direction in an equilibrating parton gas can
be written as\cite{xu96},

\begin{equation}
S(P_T) = \frac{
\int d^2r (R^2_A - r^2) exp[-\int_{\tau_{{iso}}}^{min(\tau_\psi,\tau_f)}
d\tau n_g(\tau) <v_{rel} \sigma>]}
{\int d^2 r (R^2_A - r^2)}
\end{equation}

\noindent  where $n_g(\tau)$ is the gluon density at time  $\tau$,
$<v_{rel}\sigma>$ is the thermal  averaged  gluon-$J/\psi$  cross
section,  expression  for  which  can  be  found  in \cite{xu96}.

Details   of   the   calculation   of   $n_g$  can  be  found  in
ref.\cite{ch00}. In brief, if the dominant reactions  leading  to
chemical  equilibration  process  are $gg\leftrightarrow ggg$ and
$gg\leftrightarrow   q\bar{q}$,   following   coupled   equations
determine  the  evolution  of  temperature  and  gluon and quarks
fugacities,

\begin{mathletters} \label{1}
\begin{eqnarray}
\frac   {\dot{\lambda_g}   +b_2/a_2(\dot{\lambda_q}
+\dot{\lambda_{\bar{q}}})}        {\lambda_g         +b_2/a_2(\lambda_q
+\lambda_{\bar{q}})}+     4\frac{\dot{T}}{T}     +\frac{4}{3\tau}
-\frac{4}{3}\frac{\eta}{\tau^2}                     \frac{1}{[a_2
\lambda_g+b_2(\lambda_q+\lambda_{\bar{q}})]T^4}    =&&0   \\
\frac{\dot{\lambda}_g}    {\lambda_g}   +   3
\frac{\dot{T}}{T}+\frac{1}{\tau} - R_3 (1-\lambda_g)  +  2  R_2
(1-\frac{\lambda_q       \lambda_{\bar{q}}}      {\lambda_g^2})=&&0
\label{10a}\\
\frac{\dot{\lambda}_q}          {\lambda_q}          +          3
\frac{\dot{T}}{T}+\frac{1}{\tau}    -    R_2    \frac{a_1}{b_1}
(\frac{\lambda_g}{\lambda_q}-\frac{\lambda_{\bar{q}}}
{\lambda_g})=&&0
\end{eqnarray}
\end{mathletters}

In  eqs.\ref{1},  $\lambda_g$ and $\lambda_q$  are  the gluon and quark
fugacities, defined as

\begin{mathletters}
\begin{eqnarray}
n_g(\tau) = \tilde{n}_g \lambda_g\\
n_q(\tau) = \tilde{n}_q \lambda_q
\end{eqnarray}
\end{mathletters}

\noindent where  $\tilde{n_i}$  is  the  equilibrium  density  for
parton species $i$,

\begin{mathletters}
\begin{eqnarray}
\tilde{n}_g =&& \frac{16}{\pi^2} \zeta(3) T^3 = a_1T^3\\
\tilde{n}_q =&& \frac{9}{2\pi^2} \zeta(3) N_f T^3=b_1T^3
\end{eqnarray}
\end{mathletters}

\noindent $a_2,b_2$  are  parameters  of  the  equation  of
state.  For  a  partially  equilibrating  plasma, the equation of
state was written as \cite{ch00,bi93},

\begin{equation} \label{4}
\varepsilon                          =                         3P
=[a_2\lambda_g+b_2(\lambda_q+\lambda_{\tilde q})]T^4
\end{equation}

\noindent  which  implies  a  speed of sound $c_s=1/\sqrt{3}$. In
eq.(\ref{4}), $a_2=8\pi^2/15, b_2=7\pi^2N_f/40,N_f  \approx  2.5$
is  the  dynamical quark flavors. $R_2$ and $R_3$ are the density
and velocity weighted reaction rates

\begin{equation}
R_2=1/2<\sigma_{gg \rightarrow q\bar{q}}>n_g, \hspace{.5cm}
R_3=1/2<\sigma_{gg \rightarrow ggg}>n_g
\end{equation}

\noindent  can  be  found  in  \cite{ch00,bi93}.  $\eta$  is  the
viscosity coefficient. As  it  was  done  in  our  earlier  study
\cite{ch00},  to  demonstrate  the  effect  of viscosity, we have
considered two viscosity coefficients,

\begin{mathletters}
\begin{eqnarray}
\eta_1 =&&\lambda_g \eta_g +\lambda_q \eta_q\\
\eta_2 =&& \frac{12.8}{30} \frac{a_1 T^3} {R_3/T +R_2/T}
\end{eqnarray}
\end{mathletters}

The  parameters $\eta_{g,q}$ are given in ref.\cite{ch00}. It was
seen that  while  the  temperature  evolution  and  the  chemical
equilibration rate differ considerably for viscosity coefficients
$\eta_{1,2}$,  time  integrated  signals e.g. photon and dilepton
emission rates from the equilibrating plasma are nearly same  for
$\eta_1$ and $\eta_2$ \cite{ch00}.

Initial  conditions  for  hydrodynamical  evolution are listed in
table 1. They are the result of the HIJING model calculation  for
Au+Au  collision.  HIJING  is  a  QCD  motivated phenomenological
model, as only initial direct parton scatterings are  taken  into
account.  Thus  there are some uncertainties in these parameters.
However, they suffice our purpose of demonstrating the effect  of
viscosity on gluonic dissociation of $J/\psi$.

In  fig.1,  we  have shown the $J/\psi$ survival probability from
gluonic  dissociation  at  RHIC  and  LHC  energy   for   Au+Au
collisions.  Survival probability for the ideal plasma and for the viscous
plasma,  for  the  two  viscosity   coefficients   (appropriately
labeled)  are  shown.  As  expected, survival probability at RHIC
energy is large  compared  to  LHC  energy  \cite{xu96}.  Results
confirms  our  expectation  that gluonic dissociation of $J/\psi$
increases in a viscous plasma.  Low  $p_T$  $J/\psi$'s  are  more
suppressed  than  the high $p_T$ ones. The effect of viscosity is
not large at RHIC energy. Due to viscosity  survival  probability
decreases  by  7-9\%  for  low  $p_T$  $j/\psi$'s. For high $p_T$
$J/\psi$, the  effect  is  still  lower.  Viscosity  affects  the
survival  probability of low $p_T$ $J/\psi$'s considerably at LHC
energy. For low $p_T$  $J/\psi$'s  suppression  is  increased  by
20-27\%.  As  for  RHIC  energy, effect is minimal for high $p_T$
$J/\psi$'s. $p_T$ dependence of the survival probability is easily
understood. With viscosity, pre-equilibrium stage  is  increased,
thus  low $P_T$ $J/\psi$ gets more time to interact and get lost.
Larger effect of viscosity at LHC energy is also  understood.  At
LHC  energy,  initial  parton density is high due to high initial
temperature. Pre-equilibrium stage is also longer. These leads to
more suppression at LHC energy. We also note  that  the  survival
probability  is  nearly same for both the viscosity coefficients.
Even at LHC, where the effect is most prominent,  the  difference
between  the two survival probabilities is less than 10\%. This is
in accordance with our earlier findings on  photon  and  dilepton
emission from equilibrating plasma \cite{ch00}.

\begin{figure} 
\centerline{\psfig{figure=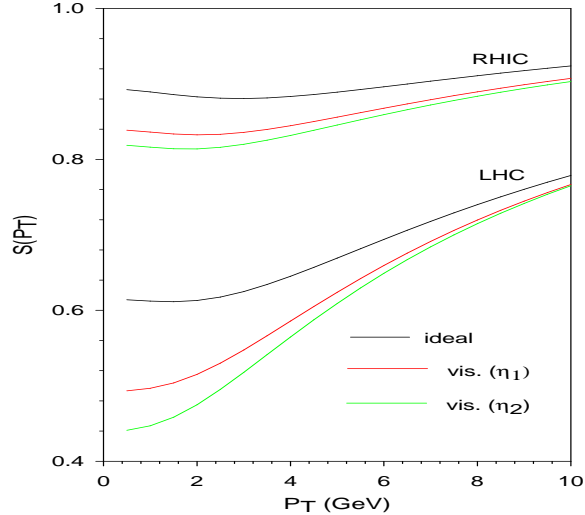,height=7cm,width=8cm}}
\caption{Survival  probability  of  $J/\psi$  in  an equilibrating
parton   plasma   at  RHIC  and  LHC  energy.}
 \end{figure}
To  summarise,  we  have  studied  the effect of viscosity on the
gluonic dissociation of $J/\psi$. Using HIJING  inspired  initial
conditions,  it  was  shown that at RHIC and LHC energies, suppression
 of $J/\psi$ due  to  gluonic  dissociation  increases
with  viscosity. Low $p_T$ $J/\psi$'s are now more suppressed.  
The effect of viscosity is not large at RHIC energy but it is 
considerable (20-30\%) at LHC 
energy. It
is hoped that results will help to have a better understanding of
$J/\psi$ suppression.

\begin{table}  \text{Initial conditions characterising the parton
plasma    at    the    onset    of    hydrodynamic     evolution}
\begin{tabular}{ccc}  &RHIC&LHC\\ \tableline $\tau_{iso}$(fm/c) &
0.31 & 0.23\\ $T_0$(GeV) & 0.57 & 0.83\\  $\lambda_g^0$&  0.09  &
0.14\\ $\lambda_q^0$ & 0.02 & 0.03 \end{tabular} \end{table}

\end{document}